# Mapping Eye Vergence Angle to the Depth of Real and Virtual Objects as an Objective Measure of Depth Perception

**Mohammed Safayet Arefin**

Department of Computer Science, Colorado State University, Fort Collins, USA.

US DEVCOM Army Research Laboratory, Los Angeles, USA, arefin@acm.org

**J. Edward Swan II**

Department of Computer Science and Engineering, Mississippi State University, Mississippi State, USA, swan@acm.org

**Russell Cohen Hoffing**

US DEVCOM Army Research Laboratory, Los Angeles, USA, russell.a.cohenhoffing.civ@army.mil

**Steven Thurman**

US DEVCOM Army Research Laboratory, Los Angeles, USA, steven.m.thurman3.civ@army.mil

# Abstract

Recently, extended reality (XR) displays including augmented reality (AR) and virtual reality (VR) have integrated eye tracking capabilities, which could enable novel ways of interacting with XR content. The vergence angle of the eyes constantly changes according to the distance of fixated objects. Here we measured vergence angle for eye fixations on real and simulated target objects in three different environments: real objects in the real-world (real), virtual objects in the real-world (AR), and virtual objects in the virtual world (VR) using gaze data from an eye-tracking device. In a repeated-measures design with 13 participants, Gaze-measured Vergence Angle (GVA) was measured while participants fixated on targets at varying distances. As expected, results showed a significant main effect of target depth such that increasing GVA was associated with closer targets. However, there were consistent individual differences in baseline GVA. When these individual differences were controlled for, there was a small but statistically-significant main effect of environment (real, AR, VR). Importantly, GVA was stable with respect to the starting depth of previously fixated targets and invariant to directionality (convergence vs. divergence). In addition, GVA proved to be a more veridical depth estimate than subjective depth judgements.

# Introduction

Over the past several decades, Extended Reality (XR) technology has progressed rapidly, leading to increasingly capable and lower-cost head-mounted displays. These displays generally encompass Virtual Reality (VR), which immerses observers in a virtual world, and Augmented Reality (AR), in which observers see virtual objects superimposed on the real world. Due to the rapid development of this technology, many novel applications of XR have been developed in areas including surgery, military training, maintenance, education, entertainment, and others (Billinghurst et al., 2015; Billinghurst & Duenser, 2012; Friedrich, 2002; Van Krevelen & Poelman, 2010). Recently, to provide automated calibration and a better user experience, many such displays have included built-in eye tracking technology (e.g., Microsoft HoloLens 2, Magic Leap 2, HTV Vive Pro Eye, and others). In this research, we measured *eye vergence angle (EVA)* from an eye-tracker and investigated how EVA varies in real, AR, and VR environments for real and simulated objects placed at different depth locations from the observer.

In XR environments, information can be presented from the observer's eye position at different depths. The accuracy and precision of depth perception of objects depends on the user's eye movements (e.g., saccadic, fixation, vergence, smooth pursuit, etc.) and the presence of monocular and binocular depth cues (e.g., accommodation, vergence, relative size, blur, binocular disparity, etc.) in the perceived scene (Arefin, Phillips, et al., 2022; Balaban et al., 2018; Singh, Ellis, & Swan, 2018). When fixating an object at a particular depth, the eyes' ciliary muscles change the eye lens shape to see information in sharp focus, known as *accommodation*, this change seeks to minimize *retinal blur* (Cholewiak et al., 2018). In addition, in binocular vision, fixating an object at a particular depth requires simultaneous *vergence eye movements*, controlled by the eye's rectus muscles. Eye vergence is primarily stimulated by *fusional vergence (disparity vergence)*: the amount of eye rotation needed to avoid double vision. Vergence eye movements bring two eye images of the viewed object to the center of the fovea (Krishnan & Stark, 1977). In addition, accommodation and vergence are coupled by the *accommodation-vergence reflex* (Hoffman et al., 2008; MacKenzie et al., 2010; Singh, Ellis, & Swan, 2018). This reflex leads to any changes in accommodation bringing about changes in vergence (*accommodative vergence*), and any changes in vergence bringing about changes in accommodation (*vergence accommodation*). Both accommodation and vergence are also linked to pupil diameter, which controls the eye's focal depth of field; the three responses all drive and are driven by each other (Myers & Stark, 1990). Under real-world binocular viewing, accommodation and vergence co-vary with the depth of the fixated object. However, when using a stereo display with a single or fixed focal plane the human visual system can override the accommodation-vergence reflex (Balaban et al., 2018; Singh, Ellis, & Swan, 2018). Because the optical depth of the display plane sets the accommodative demand, viewers using head-mounted XR displays must binocularly fuse virtual objects at depths that may differ from the display's optical depth, with a vergence demand different from the appropriate accommodation demand (Mon-Williams & Wann, 1998; Wann et al., 1995). This inconsistency is called the *accommodation-vergence* mismatch problem. This is a longstanding issue for commercial XR displays, and can cause many perceptual problems, including misperception of depth, eye strain, double-vision, and others (Hoffman et al., 2008; Kramida, 2016).

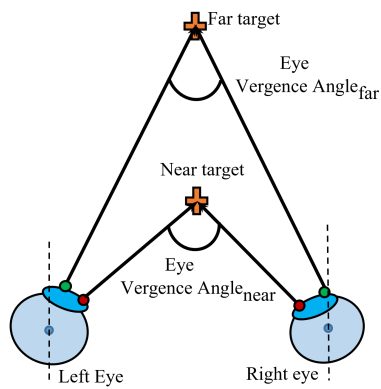

Figure 1: The geometry of the *Eye Vergence Angle* (EVA) while fixating on a near and far target. A closer target is associated with a larger EVA, and a farther target is associated with a smaller EVA.

When fixating objects at different depths, *vergence eye movements* rotate the eyes horizontally (Gross et al., 2008; Krishnan & Stark, 1977). The visual axes form the *eye vergence angle* (EVA) (Figure 1). Vergence eye movements can occur in two possible directions. When eye gaze shifts from fixating a far to a near target, the rectus muscles rotate both eyes inward, resulting in *convergence*, and when eye gaze shifts from fixating on a near to a far target, the rectus muscles rotate both eyes outward, resulting in *divergence* (Figure 1). Therefore, the value of EVA varies when fixating objects at different depths, and the geometry of binocular vision suggests that EVA is smaller while fixating on a far object, and larger while fixating on a near object. When fixating objects close to infinity, the visual axes are close to parallel, and EVA approaches zero.

In the studies of the human visual system, researchers have long used eye trackers to measure EVA. Typically, EVA has been calculated from each eye's line of sight. Primarily, EVA has been collected from fixating real objects, using controlled setups that have sometimes restricted head movements (Balaban et al., 2018; Feil et al., 2017; Hooge et al., 2019; Solé Puig et al., 2021; Sulutvedt et al., 2018; Yaramothu et al., 2018). In the XR research community, only a few prior studies have measured EVA while fixating on virtual objects, using both custom-mounted eye trackers and newer displays with built-in eye trackers. These studies have tracked perceptual depth in VR (Arefin, Swan, et al., 2022; Duchowski et al., 2022), developed a vergence-controlled gaze interaction method (Wang et al., 2022), and estimated gaze depth for varifocal displays (Dunn, 2019). Duchowski et al. (2011, 2014) directly measured and compared gaze depth between real and virtual objects. They used a commercial binocular eye tracker, a custom-built VR display, and real objects. Interestingly, they observed that vergence error was relatively small in the virtual environment compared to the physical environment. Later, Iskander et al. (2019) compared the vergence angle between ideal and VR conditions, where the EVA for the ideal condition was computed through a biomechanical simulation using inverse kinematics, and the EVA for the VR condition was computed from the gaze vectors provided by an eye tracker. They found that the EVA from the eye tracker exhibited more variability and higher values than the simulated ideal condition, which they attributed to the vergence-accommodation conflict in the VR headset.

When measuring perceived depth in XR environments, researchers have used both cognitive and perception-action based depth judgment tasks, including verbal reporting (Gagnon et al., 2021), perceptual matching (Bingham & Pagano, 1998; Swan et al., 2015), blind reaching (Swan et al., 2015), blind walking (Sahm et al., 2005), and triangulated walking (Fukusima et al., 1997). Compared to the depth of real objects in the real world, this work has generally found that the depth of virtual objects is misperceived. In AR, both depth underestimation (Gagnon et al., 2021; Swan et al., 2007, 2015) and overestimation (Livingston et al., 2009) have been reported. In VR, most previous research has found depth underestimation of VR objects compared to real objects (Interrante et al., 2006; Knapp & Loomis, 2004; Sahm et al., 2005), although, in modern VR displays, the amount of underestimation has declined (Creem-Regehr et al., 2015; Kelly, 2022; Kelly et al., 2018). Although this research has extensively explored the cognitive and perceptual phenomena behind XR depth perception, we have not found a previous study that measures EVA while fixating on objects at different depths in real, AR, and VR environments. Therefore, the effect of the environment on EVA has not previously been measured.

In this work, our initial motivation was to replicate and extend our recent work that measured EVA when fixating virtual targets in a VR environment (Arefin et al., 2022). In particular, our goal was to compare these measurements against a control condition of carefully-calibrated real targets in the real world. However, in a head-mounted VR display it is not possible to see the real world, and therefore very difficult to calibrate virtual targets to real-world targets (e.g., Steinicke et al., 2009). We solved this problem by using an optical see-through AR display (a Microsoft HoloLens 2), because in such a display, the real world can be seen, and therefore the alignment between virtual and real objects can be calibrated very precisely. Then, by covering the front of the display with an opaque cover, the HoloLens 2 can effectively become a type of VR display in which only virtual content is perceived. Therefore, this experiment used targets that were carefully matched in position and disparity depth to measure EVA in real, AR, and VR environments. All measurements were collected in a single repeated-measures experiment. To the best of our knowledge, this is the first experiment to measure vergence angle in response to carefully-calibrated objects located at different depths in real, AR, and VR environments. In addition, subjective verbal reports of perceived depth were also collected in this study to replicate previous findings of depth underestimation for XR virtual objects and for direct comparison to EVA, a more objective measure of depth perception.

Because we measured the EVA using only the gaze parameters of the eye tracker, we refer to our calculated EVA as *Gaze-measured Vergence Angle (GVA)*. In this experiment, we hypothesized that (**H1**) GVA would co-vary with the depth of fixated objects in all environments, such that fixations to nearer objects would be associated with a larger GVA and fixations to farther objects with a smaller GVA (as illustrated in Figure 1). However, due to the vergence-accommodation conflict and prior work suggesting perceptual underestimation of virtual objects, we also hypothesized that (**H2**) the functional mapping of GVA to depth in AR and VR environments would differ from the real world environment. In addition, when fixating on an object, we hypothesized that (**H3**) the mapping of GVA to object depth would not vary with the depth location of the previously fixated object. We termed this the *vergence stability hypothesis*. It predicts that, for example, when an observer fixates a target at 1.5 meters, the GVA should be the same regardless of whether the observer previously changed their gaze from .25 to 1.5 meters (divergence) or from 4 to 1.5 meters (convergence). Testing this hypothesis requires examining whether the relationship between GVA and object depth is stable across a range of convergence and divergence eye movements. Lastly, we compared GVA measurements to subjective judgments of depth perception collected from verbal reports of the depth from the same set of real, AR, and VR targets. While the finding that the depth of virtual objects is generally underestimated has been widely reported, we hypothesized that **(H4)** GVA would provide a more veridical estimate of depth for virtual objects. We reasoned that objective measurements of GVA would be less susceptible to cognitive or perceptual biases and would reflect the reflexive oculomotor response to real and virtual objects in different depth locations. Given the novelty of measuring GVA in XR environments, H4 did not predict a particular direction (i.e., overestimation or underestimation) of depth-related errors associated with GVA.

In summary, the hypotheses that motivated this experiment are:

- **H1**: Eye Vergence Angle (EVA) will covary with fixated target depth.

- **H2**: EVA will differ according to whether targets are in a real, AR, or VR environment.

- **H3**: When fixating a target, the EVA will be *stable*, regardless of the focal switching depth or vergence direction of the eye movement (*vergence stability hypothesis*).

- **H4**: EVA will provide a more veridical depth estimate than subjective depth judgments, whereas subjective judgements will be underestimated.

# Methods

## Participants

Sixteen participants, 13 males and 3 females, were recruited from the Mississippi State University community. The participants' ages ranged from 19 to 57 years, with a mean age of 29.9. Among them, 14 were right-eye dominant, and two were left-eye dominant. There were no visual restrictions for participation; 8 participants used corrective eyeglasses, while the remainder reported normal uncorrected vision. Participants' mean interpupillary distance, measured at infinity, was 64.8 mm. Each participant was compensated at a rate of $12 per hour. The Institutional Review Board of Mississippi State University approved the study protocol in accordance with the Declaration of Helsinki. Based on the data processing and quality analyses discussed below, data from 13 of these 16 participants were included in the final analysis.

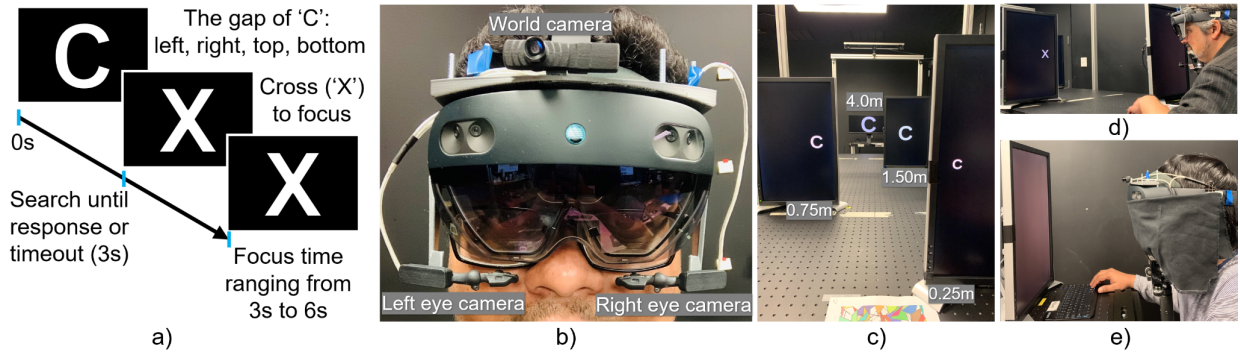

Figure 2: (a) Four-alternative forced choice visual discrimination task. (b) Pupil Labs eye tracker custom mounted on a Microsoft HoloLens 2 display. (c) Real stimuli presented on monitors at four different depths. Virtual stimuli were presented at the same depth and position for the AR and VR environments. (d) and (e) Participants performing the experiment in real and VR environments.

## Experimental Task

The experimental task aimed to obtain quantitative gaze-measured vergence angle (GVA) measurements as participants shifted their gaze to fixate on stimuli at different depths. A four-alternative forced choice visual discrimination task ensured that participants successfully shifted their fixation from one target to the next. The Landolt C task was employed. A trial began when a single capital letter "C" in a sans-serif font (Arial) appeared at a certain depth. For each trial, the gap of the "C" was randomly chosen to be left, right, top, or bottom. The participants' task was to determine the gap direction, and to provide this response within 3 sec. After the participant's response, or after the maximum allowed time of 3 sec, the letter "C" changed to an "X" (Figure 2a). Participants were instructed to focus on the center of the "X" for a variable inter-trial interval that lasted for a randomly-chosen time interval of 3 to 6 sec. When this time interval was finished, the trial ended. Next, a new letter "C" appeared at a new depth location, beginning the next trial. Participants were instructed to shift their gaze promptly to the new target letter.

## Apparatus and Setup

The real visual stimulus in the real environment was presented on four physical monitors positioned at four different depths relative to the participant's eye location (Figure 2c). Three of the monitors (Dell U2211H) were identical, each with a display resolution of 1920×1080, 40.05 pixels per cm, a diagonal size of 55 cm, and placed on the top of a 244 cm by 92 cm optical breadboard. The fourth monitor (Dell Ultrasharp Monitor U2913WM) had a display resolution of 2560×1080, 73.66 pixels per inch, a diagonal size of 73.66 cm, and was placed on the top of a locked rolling cart. The four monitors were positioned in such a way that participants had a clear line of sight to the visual targets presented on each monitor. The size of the target stimulus at each distance was scaled to achieve a constant size in terms of degrees of visual angle relative to the observer.

Throughout the experiment, participants wore a Microsoft HoloLens 2 binocular see-through head-mounted display (Figure 2). According to the manufacturer (Microsoft), the Hololens 2 has a 2K resolution, weighs 566 grams, and has a diagonal field of view of 52°. Participants wore the display when observing real target objects (Figure 2d), but it was switched off. Participants saw target objects on the four monitors. When observing AR target objects, participants saw virtual target objects positioned in the experimental room. When observing VR target objects (Figure 2e), the display was covered with a black cloth, completely blocking the experimental room's view. Participants saw virtual target objects floating in blackness. The cloth was carefully positioned not to block the HoloLens 2 tracking cameras, so the display's tracking algorithm ran as normal. The stimulus at the far distance was observed perpendicularly at the participant's midline. Stimuli at other distances were positioned in such a way as to require head rotations of 25° right, 7.60° left, and 5.71° right, respectively, in order to shift fixation from perpendicularly facing the far distance to perpendicularly facing the new distance. However, the experimenters observed that the actual amount of head rotation varied slightly between different participants. Both real and virtual stimuli were rendered in white, and each stimulus had a constant visual angle size of 3.87°. This ensured that the stimuli size did not vary with distance, similar to previous experiments conducted in this laboratory (Arefin et al., 2020; Arefin et al., 2022).

When using optical see-through AR displays, placing virtual objects at very precise locations in the real world has historically been challenging. We used a method that we developed for a previous set of experiments (Khan et al., 2021a, 2021b, 2022), which uses an optical tracking algorithm that operates through the HoloLens' forward-facing camera to establish the virtual location of a real world fiducial mark. Previous experimentation showed that with this approach, the perceived alignment error between real and virtual object locations could be as low as a few millimeters (Khan et al., 2022).

Calculating GVA requires an eye gaze ray (origin and direction) from both eyes. The Microsoft HoloLens 2 has eye-tracking cameras for both eyes, but when this experiment was conducted, Microsoft's MixedReality Toolkit (MRTK) eye-tracking API only provided a single eye gaze ray[1]. It was not possible to know whether the right or left eye gaze ray was being returned, and it was generally not possible to compute GVA from the HoloLens 2 eye tracking cameras. Therefore, a Pupil Labs Pupil Core stereo eye tracking system was mounted to the HoloLens 2, using a custom-designed 3D-printed scaffolding (Figure 2b). Each eye tracking camera was carefully positioned to have a clear view of the eye, and not to block any of the participant's field of view. According to the manufacturer (Pupil Labs), the eye tracking system has an accuracy of .60°, a precision of .02°, a sampling rate of 200 Hz, and a resolution of 192x192.

---

[1]More recently, the MRTK eye-tracking API has added the ability to obtain ray origin and direction from both built-in eye tracking cameras.

The entire experiment was controlled by two separate Unity programs, a stand-alone Unity application (Unity version: 2021.3.4f1) for real targets, and another mixed reality Unity application (Unity version: 2020.2.2f1) for AR and VR targets.  Both programs used Microsoft's MixedReality Toolkit (MRTK), and ran on a Windows 10 desktop (HP Z2 Tower G4 Workstation), with an Intel Core i-9 CPU running at 3.60GHz with 64GB of RAM. The eye tracking data was gathered using the Pupil Labs eye tracker API for Unity. The offline eye movement data processing, analysis, and visualizations were performed with Matlab (R2022a) and R (4.2.3). A numeric keyboard was used to gather the participant's responses.

## Variables and Design

The experiment examined two different independent variables: *environment* and *depth*.

Three levels of the *environment* were considered: *real, AR,* and *VR*. Under the real condition, participants observed the real stimuli on physical monitors (Figures 2c and 2d). For the AR and VR conditions, participants observed the stimuli through the HoloLens 2 display. For the VR condition, the display was covered with a black cloth that occluded the view of the real world (Figure 2e). To control for the effect of wearing the head-mounted display and equipment, participants wore the HoloLens 2 display for every level of environment, including in the real condition.

The distance from the participant's eye position to the stimulus position is *depth*. Four levels were considered (Table 1): 0.25m (4.0D), 0.75m (1.33D), 1.50m (0.67D), and 4.0m (0.25D). In each experimental trial, participants made a *vergence eye movement* that changed their gaze from a target at the *start depth* to a target at the *end depth*, resulting in a particular *focal switching depth*. These depth changes resulted in 12 possible permutations; 6 were divergence eye movements, and 6 were convergence eye movements (Table 1).

Table 1: The experiment placed targets at 4 different depths, resulting in 12 different vergence eye movements.

| Depth Pair | Start Depth | End Depth | Focal Switching Depth (\|Start Depth - End Depth\|) | Vergence Eye Movement |
|---|---|---|---|---|
| 1 | 0.25m (4.0 D) | 0.75m (1.33D) | 0.5 m (2.67D) | Diverge |
| 2 | 0.25m (4.0 D) | 1.50m (0.67D) | 1.25m (3.33D) | Diverge |
| 3 | 0.25m (4.0 D) | 4.0 m (0.25D) | 3.75m (3.75D) | Diverge |
| 4 | 0.75m (1.33D) | 1.50m (0.67D) | 0.75m (0.66D) | Diverge |
| 5 | 0.75m (1.33D) | 4.0 m (0.25D) | 3.25m (1.1 D) | Diverge |
| 6 | 1.50m (0.67D) | 4.0 m (0.25D) | 2.50m (0.42D) | Diverge |
| 7 | 4.0 m (0.25D) | 1.50m (0.67D) | 2.50m (0.42D) | Converge |
| 8 | 4.0 m (0.25D) | 0.75m (1.33D) | 3.25m (1.1 D) | Converge |

| 9  | 4.0 m (0.25D)  | 0.25m (4.0 D)  | 3.75m (3.75D)  | Converge |
| 10 | 1.50m (0.67D)  | 0.75m (1.33D)  | 0.75m (0.66D)  | Converge |
| 11 | 1.50m (0.67D)  | 0.25m (4.0 D)  | 1.25m (3.33D)  | Converge |
| 12 | 0.75m (1.33D)  | 0.25m (4.0 D)  | 0.5 m (2.67D)  | Converge |

A within-subjects, repeated-measures design was employed. Each combination of environment and depth pair was repeated six times in pseudorandom order. Therefore, each participant observed 3 (*environment*) × 12 (*depth pair*) = 36 conditions, where each condition was repeated 6 times, for a total of 216 trials per participant in a complete experimental session. Between participants, the presentation order of *environment* was randomly chosen from either *real-AR-VR*, *AR-VR-real*, or *VR-AR-real*. Within each participant, the presentation order of *depth pair* × *repetition* was randomly permuted, with the restriction that there was no back-to-back repetition of the same distance. At the beginning of the experiment, the first *start depth* was chosen randomly. For each subsequent trial, the end depth of trial *i* was the start depth of trial *i* + 1.

For each trial, the participant's continuous real-time left and right eye movement data (gaze direction vector, gaze origin vector, gaze normal vector, eye center vector, gaze data confidence level), the response of the Landolt C task (left, right, bottom, top), and the button press timestamp were collected. For each participant, 3 verbal distance estimates were also collected for each environment (real, AR, VR) and distance (.25, .75, 1.5, 4 meters).

## Procedure

The experiment was conducted in a 7.68 by 5.46 meter room. As the room was generally used for optical experiments, it was painted black and had no windows. The room contained a large 244 cm by 92 cm optical breadboard on which experimental equipment was positioned. During the experiment, participants were seated at one of the narrow ends of the optical breadboard (Figure 2).

A participant began the experiment by signing a consent form and filling out a general experimental questionnaire. Next, the participant's interpupillary distance was measured at infinity with a commercial pupilometer, and Miles's test (1930) was employed to determine the dominant eye. Then, a detailed description of the experiment was provided. During this process, the experimenter presented examples of stimuli on paper so the participant could familiarize themselves with the stimuli and the task. Next, the experimenter helped the participant put on the HoloLens 2, seeking a proper fit that reduced unwanted slippage.

Upon donning the HoloLens 2, the participant performed two separate calibration procedures. First, the participant performed the standard HoloLens 2 calibration procedure. This ran the proprietary HoloLens 2 calibration software, which used the built-in HoloLens 2 eye trackers. We can assume that the calibration measured the participant's interpupillary distance and other features of the eye position relative to the display's tracking coordinate frame (Grubert et al., 2017).

Second, the participant performed a procedure to calibrate the Pupil Labs Pupil Core eye tracker. The participants sat on a chair within arm's length distance (~60 cm) from the experimenter's monitor. The experimenter then adjusted the eye tracking cameras so the participant's pupils and eyeballs were clearly visible and checked the pupil labs' pupil capture software to ensure that the pupil capture model was well-fitted to both eyes. Then, the calibration procedure was run. Five circular markers were displayed: four markers on the four corners and one at the center. Participants were instructed to follow and focus on the center of the calibration marker with minimal head movement.

After properly performing the display and eye tracker calibrations, participants sat on a tall height-adjustable chair, placed their chin on a chin rest, and placed their right hand on the numeric keyboard. The chair height was adjusted so that participants of different heights could sit comfortably and all targets were at the same height as the eyes. The chin rest minimized head movement during the experiment.

For VR and AR environments, participants next performed a third calibration procedure. This calibration, adopted from previous experiments in our lab (Khan et al., 2021a, 2021b, 2022), allowed the HoloLens 2 to position virtual objects with respect to the real world precisely. Participants were instructed to look at a printed fiducial marker that was attached to the optical workbench (visible at the bottom of Figure 2c). A machine vision tracking algorithm (Vuforia, PTC Inc.) running through the HoloLens forward-facing camera recognized the marker, and aligned the HoloLens virtual coordinate system with respect to the marker's location. The HoloLens then rendered a virtual cube. Participants reported if the virtual cube was located in the middle of the fiducial marker, and properly oriented. If so, the virtual coordinate system was properly calibrated. If not, the tracking system was restarted, and the procedure was repeated.

Next, participants were shown the Landolt C target object, with the gap located to the right, at all four distances at the same time (Figure 2c). Participants were asked to verbally estimate the distance to each target, starting from the farthest and moving closer. Participants could use their desired units (feet, meters, inches). The experimenter encouraged the participant to modify their estimate as desired, and continued prompting until the participant had provided 3 estimates of each distance. Many, but not all, participants chose to give the same value 3 times. The experimenter recorded these verbal responses on paper.

Next, the main experimental trials for the current environment were presented. After the experiment, an informal interview session was performed to collect additional thoughts, impressions, and insights. The participant was then compensated. The experiment lasted approximately one hour.

## Data Pre-Processing

Before calculating GVA, we preprocessed the eye tracker data to omit blinks and noise artifacts associated with physiologically unrealistic data points. For omitted data points, we simply replaced them with "not a number" (NaN) in Matlab to preserve the full temporal structure of the raw data while ensuring that they would not be incorporated into the overall mean estimates.

We first preprocessed the collected eye tracker data by using the confidence value of each data point. According to the Pupil Labs eye tracker API, confidence values ranged continuously from 0 to 1, where 0 means the pupil was not detected at all, while 1 means the pupil was detected with high reliability. Data points with low confidence < .75 were replaced with NaN values. Next, we performed velocity-based data filtering on vergence data to discard unrealistic physiological GVA values. Under this criteria, data points with very large velocities (GVA > 5000 deg/sec) were replaced with NaN values. Then, data points ≥ 2.5 standard deviation (SD)s from the mean were replaced with NaN values. When applied to our data, these steps resulted in excluding an average, over all participants, of 16.72% of the data points for Real (min participant: 4.19%, max participant: 47.20%); 14.58% for AR (min: 2.72%, max: 43.52%); and 18.54% for VR (min: 3.65%, max: 27.08%).

After obtaining the pre-processed eye-gaze data values from the previous step, we checked trial-by-trial validity. A trial with > 50% valid eye-gaze data samples was considered valid. Based on this criteria, a total of 842 valid real trials (90.00% of the real trials), 893 valid AR trials (95.41% of the AR trials), and 879 valid VR trials (93.91% of the VR trials) were included.

The study considered 4 target distances that created 12 distinct distance pairs (Table 1). Each distance pair was repeated 6 times per participant per environment. Distance pairs that contained ≥ 3 valid trials out of 6 were considered valid. In addition, each participant experienced each environment. Environmental conditions with ≥ 6 valid distance pairs out of 12 were considered valid. A participant was considered valid if the participant's data had three valid environmental conditions. According to this criteria, the data from 2 participants were discarded from the analysis. Another participant's data was excluded due to eye tracker calibration issues. Therefore, 13 out of 16 participants were included in the analysis.

For each trial, we sought the first fixation that occurred ≥ 250 ms after stimulus onset, but before the button response was made. The analysis then focused on a time window of 1s to 2s after fixation onset (Figure 3b). This ensured that GVA had enough time to stabilize on each new target depth. We chose the 1s to 2s window for GVA analysis based on the timing of the perceptual and oculomotor system to perceive a target and make a saccadic eye movement to the correct spatial location (~250 ms), and the vergence eye movement to allow sharp focus on the stimulus (200–300 ms) (Dunn et al., 2020; Puig et al., 2013; Solé Puig et al., 2021).

## GVA Calculation

When calculating GVA , we assumed unrestricted head and eye movements. Given that the 3D gaze direction vectors from the left and right eyes came from head-mounted eye trackers, we could assume that the 3D gaze vectors included head rotations, vergence eye movements, and interpupillary distance (Dunn, 2019). However, our experimental target positions do not meet the assumptions of the standard *triangulation* method because targets are not positioned perpendicularly from the eye position. Also, in our experiment, participants are free to rotate their heads to view the targets frontally. Previously, for experiments with real target objects, many researchers calculated vergence angle from 3D gaze direction vectors and vector intersection models (Feil et al., 2017; Solé Puig et al., 2021; Sulutvedt et al., 2018). Further, for experiments with virtual target objects seen in AR and VR environments, previous researchers also calculated vergence angle from 3D gaze vectors (Arefin, Swan, et al., 2022; Dunn, 2019; Wang et al., 2022), and found that the vector intersection model was appropriate when eye movements were free (Arefin, Swan, et al., 2022; Duchowski et al., 2014). Therefore, we calculated gaze-measured vergence angle (GVA) from the angle formed by the left and right eye 3D gaze vectors (Figure 3a). The 3D gaze direction vectors ( $\boldsymbol{L}_{(x,y,z)}$ and $\boldsymbol{R}_{(x,y,z)}$) were projected on a plane, where $x$ is the horizontal, $y$ the vertical, and $z$ the optical (depth) axes in the pupil labs 3D camera space coordinate system. Then, GVA was calculated as

$$\mathrm{EVA} = \cos^{-1}\left(\frac{\boldsymbol{L}_{(x,y,z)} \bullet \boldsymbol{R}_{(x,y,z)}}{|\boldsymbol{L}_{(x,y,z)}||\boldsymbol{R}_{(x,y,z)}|}\right).$$

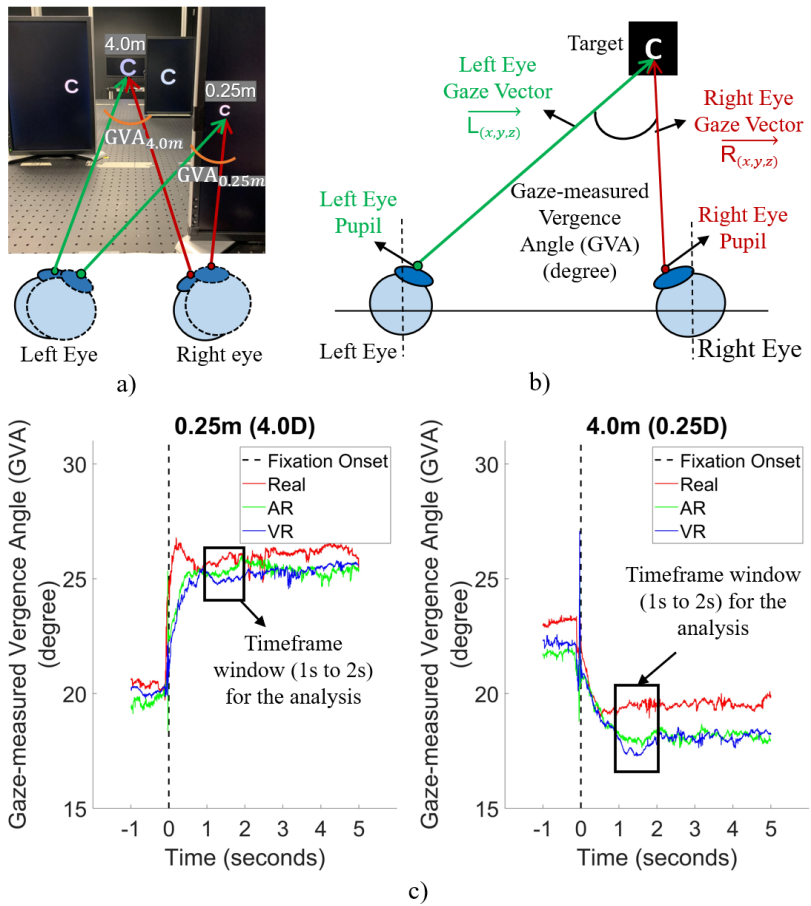

Figure 3: Gaze-measured Vergence Angle (GVA) calculation and an example of unprocessed eye tracking data. (a) Participant needs to rotate different amounts of head and eye to focus on objects at 0.25m and 4.0m. (b) GVA is calculated from the left and right eye 3D gaze direction vectors for the distance of 0.25m. (c) An example of unprocessed eye tracking data at 0.25m (4.0D) and 4.0m (0.25D) for a participant. The vertical dotted line at 0 seconds shows fixation onset, which occurs approximately 250 ms after stimulus onset, and before the Landolt C button press. This example shows the difference in GVA values for the experiment's two extreme depths (near and far) in the Real, AR and VR environments. For GVA analysis, a time window between 1 and 2 seconds after the fixation onset was used.

# Results

As discussed above, the Landolt C task required a four-alternative forced-choice visual discrimination. A high score on this task demonstrates that participants attended to the task and focused at the correct depth. Behavioral results showed that participants' mean accuracy was very high. The grand mean across all subjects and environmental conditions was 98%, with a standard deviation of 2.5% and a range of 90–100%. This indicates that all 13 participants paid attention and focused on the target at the presented depth location in nearly every trial.

Results were analyzed using multiple linear regression (Pedhazur, 1997). Multiple regression methods are sometimes preferable to standard ANOVA analysis, because they allow predicting a continuous dependent variable (*GVA*) from a combination of continuous (*eye movement end depth*) and categorical (*environment*: Real, AR, or VR) independent predictor variables. When an independent variable is inherently continuous, multiple regression methods have more power to detect significant effects than standard ANOVA analysis (Pedhazur, 1997). In addition, multiple regression yields slopes and intercepts, which are useful descriptive statistics for data of this form. Finally, multiple regression encourages statistical reasoning in terms of $R^2$, the percentage of observed variation explained by a given linear model. $R^2$ is a normalized and intuitive measure of effect size.

Each analyzed result then began with a *complete model*, which for each predictor included the main effects and all interactions. This complete model was then subjected to a stepwise refinement algorithm, where terms were dropped with the goals of (1) simplifying the model, while (2) not significantly reducing the model fit[2]. This resulted in a *fitted model*. The fitted model was validated by considering $dR^2$, the reduction in explained variance from the complete to the fitted model. Also considered was whether further reducing the model resulted in a significant $dR^2$ change from the fitted model, and whether the fitted model suggested an explanation that aligned with knowledge of eye movement behavior. Based on these considerations, the stepwise refinement technique was judged to provide statistical evidence that the fitted model was the most appropriate model to explain the results.

We also analyzed the data with mixed models, using the R *lme4* package (Bates et al., 2015), with *participant* as a random factor, and *end depth* and *environment* as fixed factors. The mixed model analysis showed that participant intercept, but not slope, was the dominant random effect. In this section, participant intercept and slope were carefully analyzed and discussed. However, we were concerned that, in the context of mixed models, there is not a general agreement on how to calculate $R^2$, and exactly what $R^2$ means (Jaeger et al., 2017). We wanted to use $R^2$ in a context where the interpretation of $R^2$ as a normalized, unitless measure of effect size is well established and well understood. Therefore, the paper uses regular linear models and reports $R^2$ for all effects.

---

[2] The R `step()` function (Hastie & Pregibon, 1992; R Core Team, 2023).

## End Depth and Environment

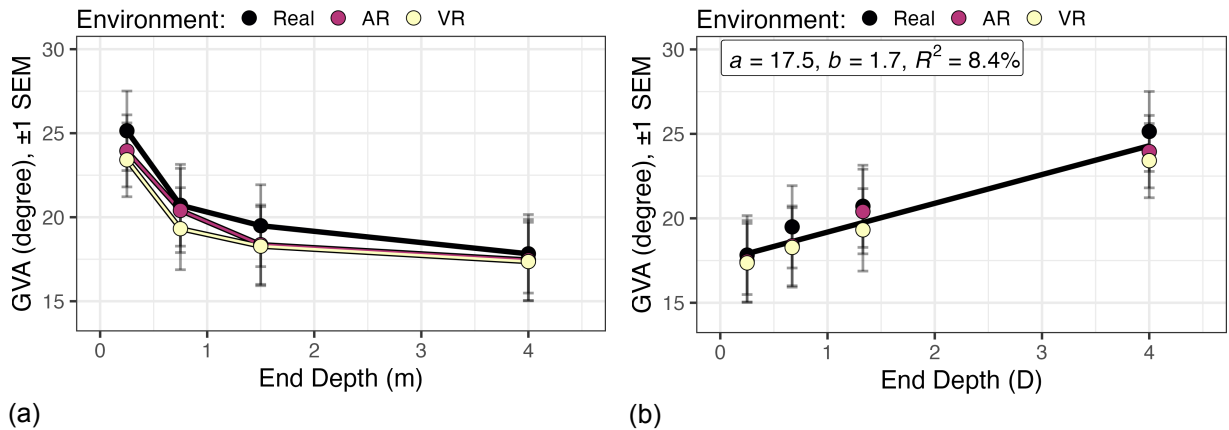

(a)                  (b)

Figure 4: The Gaze-measured Eye Vergence Angle (GVA) as predicted by *environment* and eye movement *end depth* in meters (a), and in diopters (b). In (b), the fitted linear model *GVA ~ end depth* is shown, with the parameters of intercept (*a*), slope (*b*), and percentage of explained variation ($R^2$). *N* = 156 data points. See Figure 5.

```
Model    Formula                              R2      Res.Df   Df    F       p
cm       GVA ~ Environment * EndDepth         8.76%   150
fm       GVA ~ EndDepth                       8.39%   154      -4    0.2     0.962
rm       GVA ~ 1                              0.00%   155      -1    13.8    <0.001 ***

R2 explained by:
    EndDepth =        fm/cm = 95.8%
Environment = (cm-fm)/cm =   4.2%
```

Figure 5: Multiple regression analysis from Figure 4. The fitted model is highlighted. *Model*: cm = complete model; fm = fitted model; rm = reduced model. *Formula*: The regression formulas, given in R format (R Core Team, 2023). *R2*: The percentage of variation explained by the model. *Res.Df*: The residual degrees of freedom left in the data. *Df*: The degrees of freedom used by the model. The subtraction leads to the Res.Df of the line above. *F, p*: An *F*-test of the reduction in $R^2$ between the current line and the line above (Pedhazur, 1997). *R2 explained by*: The percentage of $R^2$ explained by the predictor variables of interest.

## *GVA Results*

We hypothesized that eye vergence angle (GVA) would covary with the depth of fixated target objects (H1), and that this covariation would also differ according to whether the environment included Real, AR, or VR target objects (H2). Figure 4 shows the effects of environment and eye movement end depth on mean GVA. Figure 4a shows the depth in meters, while Figure 4b shows the same data with depth expressed in diopters (1/distance). While there is a non-linear relationship between depth and GVA due to the geometry of the eye vergence system (Figure 1), this relationship can be expressed as a linear relationship by converting depth to units of diopters representing the inverse of distance. Our statistical modeling was therefore applied to data in which depth is expressed in diopters. The dataset has 2489 vergence measurements (trials), which were averaged over participant (13), environment (3), and end depth (4), resulting in $N$ = 156 analyzed values[3]. A complete linear model was created (Figure 5) that predicts the measured GVA from the contributions of the eye movement end depth and the environment (Real, AR, VR): *GVA ~ end depth * environment*[4]. This complete model explains $R^2$ = 8.8% of the variation. As shown in Figure 4b, the fitted model is *GVA ~ end depth*, explaining $R^2$ = 8.4% of the variation, an insignificant difference of $dR^2$ = .4% ($F_{4,150}$ < 1). The fitted model explains significantly more variation ($F_{1,154}$ = 13.8, $p$ < .001) than the next reduced model, *GVA ~ 1* (the grand mean). Of the original explained variation of $R^2$ = 8.8%, 95.8% is explained by end depth, and 4.2% by environment (Figure 5). Therefore, the statistical evidence is that the end depth of the eye movement, but not the environment, significantly predicted *GVA* over this range of distances. While this analysis supports hypothesis H1, but does not support hypothesis H2, we suspected that individual differences obscured the effect of the environment.

*Variation Per Participant*

Individual differences were also analyzed. Figure 6 shows the data from Figure 4b separated by *participant*, with a linear model fitted for each participant. Note that participant intercepts (*a*: $M$ = 17.5°, $SD$ = 8.6°) have much higher variation than slopes (*b*: $M$ = 1.7$^{°/D}$, $SD$ = 0.37$^{°/D}$). A model that only predicts GVA by intercepts (*GVA ~ participant*) explains $R^2$ = 87.5% of the variation, while the model that adds slopes (*GVA ~ participant * end depth*) explains $R^2$ = 96.3%, an increase of only $dR^2$ = 8.8%. Therefore, the great majority of the variation (87.5 / 96.3 = 90.9%) shown by the error bars in Figure 4 comes from this per-participant variation in intercept as visualized in Figure 6.

---

[3]As discussed above in *Data Pre-Processing*, there were different repetition counts of vergence measurements per experimental condition.
[4] Regression formulas are given in R language format (Chambers & Hastie, 1993; R Core Team, 2023).

Participant:

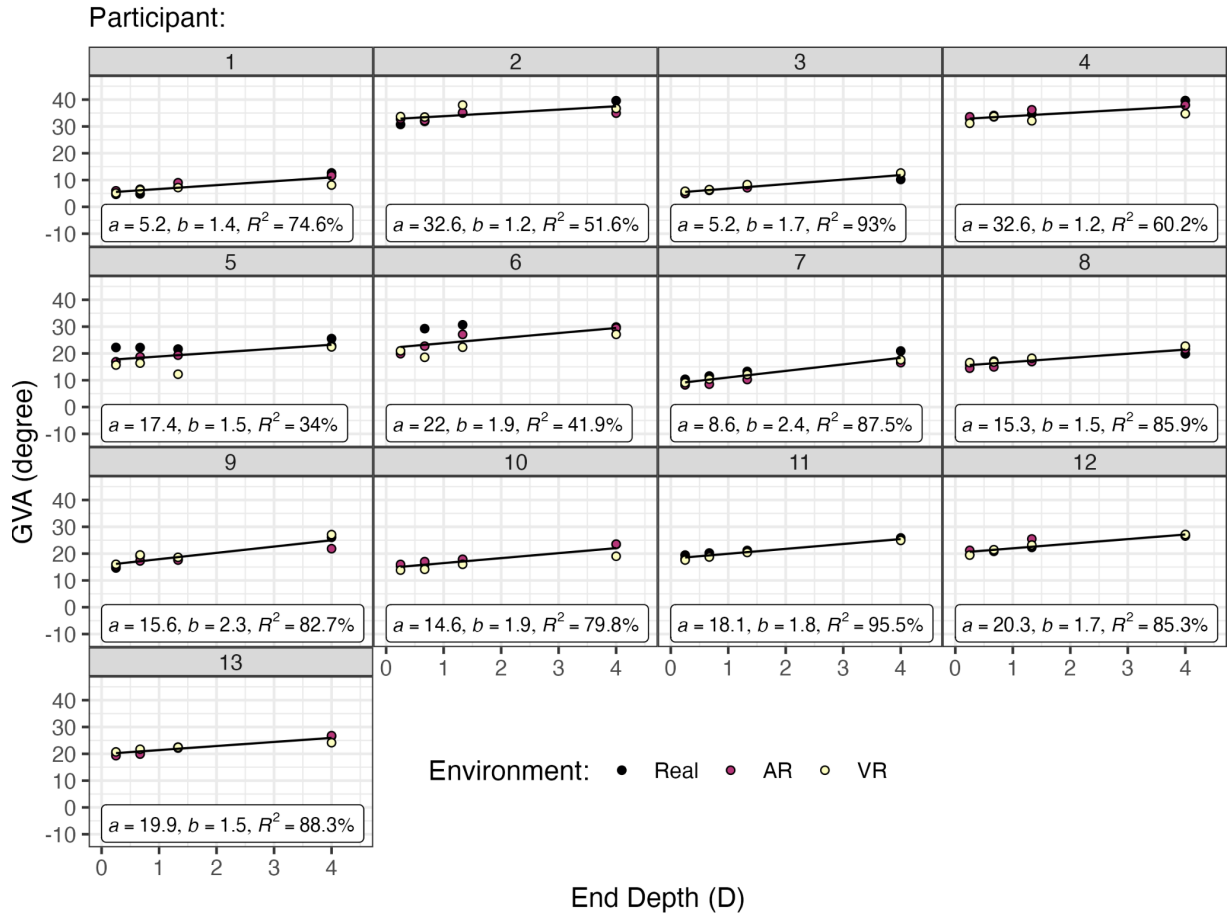

Figure 6: The GVA as predicted by eye movement *end depth* and *participant* (data from Figure 4b). The x-axis represents target depth in diopters (1/distance). The fitted model for each participant is added. Note the large variation in intercept (*a*), compared to slope (*b*). *N* = 156 data points.

## Normalized GVA Results

As discussed above, the fitted model in Figure 4b was hypothesized to support hypothesis H1, that GVA covaries with end depth, and hypothesis H2, that this covariation would differ according to environment (Real, AR, or VR). However, the fitted model (*GVA ~ end depth*) supported H1 but not H2. Nevertheless, 4.2% of the explained variation (out of a total $R^2$ = 8.8%; Figure 5) came from environment. Because the per-participant intercept variation was such a strong effect ($R^2$ = 87.5%), we suspected that this variation might be masking an environment effect that would become detectable if the experiment had more power. In other words, the lack of a significant environment effect could be a Type II error (Howell, 2001): the effect is present and reliable, but the current experiment does not have enough power to detect it. While the best way to address a lack of power would be to collect data from more participants, we decided to remove the intercept differences in the current data and see if the environment effect became detectable. In addition, we decided removing intercept differences was further justified by the idea, as discussed below, that the large per-participant variation might be due to the experimental issue of eye tracking camera alignment, which we expect to vary each time a headset is placed on the head, and not due to inherent differences in participant physiology, such as interpupillary distance.

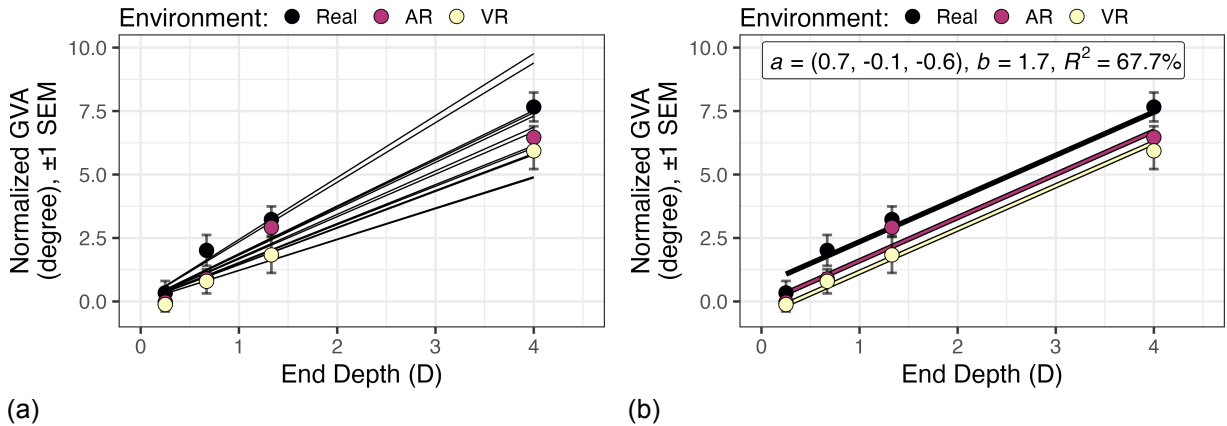

(a)                                                    (b)

Figure 7: The GVA, normalized by subtracting the intercept of each participant (Figure 6), predicted by *environment* and eye movement *end depth*. Target depth on the *x*-axis is represented in diopters (D), computed as 1/distance. (a) A linear model for each participant is added: *normalized GVA~ participant* (Figure 6). (b) The fitted linear model *normalized GVA~ environment + end depth* is shown. *N* = 156 data points. See Figure 8.

```
Model   Formula                              R2      Res.Df   Df    F      p
cm        NormGVA ~ EndDepth * Environment   67.97%   150
fm        NormGVA ~ EndDepth + Environment   67.72%   152     -2    0.6    0.56
rm        NormGVA ~ EndDepth                 65.13%   154     -2    6.1    <0.01 **

R2 explained by:
    EndDepth =       rm/fm  96.2%
Environment = (fm-rm)/fm   3.8%
```

Figure 8: Multiple regression analysis from Figure 7. The fitted model is highlighted. See the caption for Figure 5.

Therefore, *normalized GVA = GVA– per-participant intercept* (the values of *a* in Figure 6) was computed for all 2489 vergence measurements. Here, normalized GVA holds the residual values when the strong effect of per-participant intercept is removed. Figure 7a shows the same analysis as Figure 6, but is plotted against normalized GVA. The lines represent the same 13 per-participant linear models. Here, for each model intercept *a* = 0, but the values for slopes *b* and $R^2$ are the same as in Figure 6. In Figure 7, note how much smaller the error bars are compared to Figure 4; this is a visual indication of how much variation has been removed from normalized GVA, compared to GVA.

The same regression analysis was conducted as before. The complete model (Figure 8) is *normalized GVA~ end depth * environment*, explaining $R^2$ = 68.0% of the variation. As shown in Figure 7b, the fitted model is *normalized GVA~ end depth + environment*, explaining $R^2$ = 67.7% of the variation, an insignificant difference of $dR^2$ = .25% ($F_{2,150}$ < 1). The fitted model explains significantly more variation ($F_{2,152}$ = 6.1, $p$ < .01) than the next reduced model, *normalized GVA~ end depth*. Of the explained variation of $R^2$ = 67.7%, 3.8% is explained by environment, and 96.2% by end depth.

Therefore, as anticipated, for *normalized GVA*, the fitted model supports both hypothesis H1 and hypothesis H2. Here, because the strong signal of per-participant intercepts has been removed, the fitted model provides statistical evidence for an effect of environment. However, note that the effect size of environment, 3.8% of explained variance, is no larger than the corresponding value for the analysis in Figure 4b, 4.2% of explained variance. Therefore, environment has a small but statistically significant effect, which cannot be detected with a model regressed against GVA, but can be detected with a model regressed against normalized GVA.

In addition, the fitted model is additive, *end depth + environment*, meaning that main effects of end depth and environment were detected, but not an interaction between them. Therefore, the fitted model lines in Figure 7b differ only in intercept, $a$ = (.7, –.1, –.6), but not slope, $b$ = 1.7. Note that the slope is the same in Figures 7b and 4b. Therefore, measured vergence angles for AR targets were .8 degrees smaller than for Real targets, and for VR targets were 1.3 degrees smaller than for Real targets. These differences are constant across the range of end depths (.25 to 4 D) that were studied. Whether these differences are meaningful will depend on the application for which GVA is being measured.

## Vergence Stability at End Depth

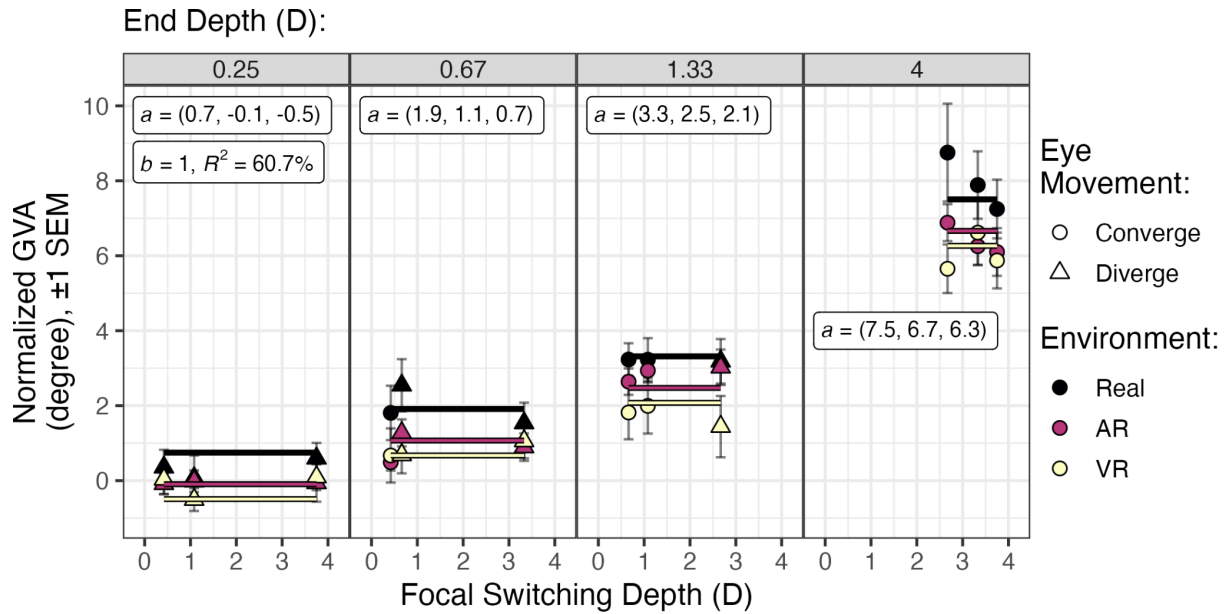

Figure 9: The normalized GVA predicted by *focal switching depth*, *environment*, and eye movement *end depth*. The direction of vergence eye movements is also indicated. Focal switching depth on the x-axis, and target depth across the panes, is represented in diopters (D), computed as 1/distance. The fitted linear model *normalized GVA~ environment + end depth* is shown. N = 465 data points. See Figure 10.

```
Model   Formula                            R2    Res.Df   Df     F      p
cm1     NormGVA ~ SwitchDepth * EndDepth * Env   61.88%   441
cm2     NormGVA ~ EndDepth * Env           61.35%   453    -12   0.5    0.907
fm      NormGVA ~ EndDepth + Env           60.66%   459    -6    1.3    0.240
rm      NormGVA ~ EndDepth                 58.28%   461    -2    13.8   <0.001 ***

R2 explained by:
    EndDepth =         rm/cm1 = 94.2%
Environment =   (cm2-rm)/cm2 =  5.0%
SwitchDepth = (cm1-cm2)/cm1 =  0.9%
```

Figure 10: Multiple regression analysis from Figure 9. The fitted model is highlighted. See the caption for Figure 5. Note that here there are two different complete models: *cm1*, which contains all of the predictor terms, and *cm2*, from which the switching depth predictor has been removed. See the caption for Figure 5.

```
Model   Formula                        R2     Res.Df   Df     F      p
cm      GVA ~ SwitchDepth * Env * EndDepth   9.60%   441
fm      GVA ~ EndDepth                 9.10%   461    -20   0.1    1
rm      GVA ~ 1                        0.00%   464    -3    14.8   <0.001 ***
```

Figure 11: The same multiple regression analysis as Figure 10, but regressed against *GVA* instead of *normalized GVA*. Note the much smaller $R^2$ values, and that the fitted model does not include

We also expected to see evidence of the *vergence stability hypothesis* (H3): when observers fixate a target, the GVA will be *stable*, regardless of the focal switching depth distance or vergence direction of the eye movement. Figure 9 shows the effects of environment, end depth, and focal switching depth on mean normalized eye vergence angle. In addition, the direction of the eye movement (converge, diverge) is also shown. For this analysis, the 2489 vergence measurements were averaged over participant (13), environment (3), and the 12 different combinations of end depth and focal switching distance (Table 1), resulting in 468 conditions. However, as discussed above there were different counts of valid vergence measurements (trials) per condition, and for this analysis, the data is missing three conditions, resulting in $N$ = 465 analyzed values. A complete linear model was created (Figure 10) that predicts normalized GVA from the contributions of the focal switching depth of the eye movement, the end depth of the eye movement, and the environment (real, AR, VR) in which the eye movement took place: *normalized GVA ~ switching depth * end depth * environment*. This complete model explains $R^2$ = 61.9% of the variation. As shown in Figure 9, the fitted model is *normalized GVA ~ environment + end depth*, suggesting that normalized GVA is significantly predicted by the additive effects of end depth and environment. This fitted model explains $R^2$ = 60.7% of the variation, an insignificant difference of $dR^2$ = 1.2% ($F_{18, 441}$ < 1). The fitted model explains significantly more variation ($F_{2, 459}$ = 13.8, $p$ < .001) than the next reduced model, *normalized GVA ~ end depth*. Of the original explained variation of $R^2$ = 61.9%, 94.2% is explained by the end depth of the eye movement (Figure 10). Notable here is just how little explanation comes from focal switching depth: 0.9%. In addition to the terms in the fitted model, this small effect size is further statistical evidence for the validity of the vergence stability hypothesis. Thus, this analysis supports hypothesis H3.

Finally, when regressed against normalized GVA, environment explains a small but detectable 5% of the explained variation of $R^2$ = 61.4% (Figure 10; note that factoring out switching depth requires a different denominator for environment). As in the analysis above, when the same complete model is regressed against GVA (instead of normalized GVA), the fitted model does not include environment: *GVA ~ end depth* (Figure 11).

## Subjective Depth Judgements and GVA

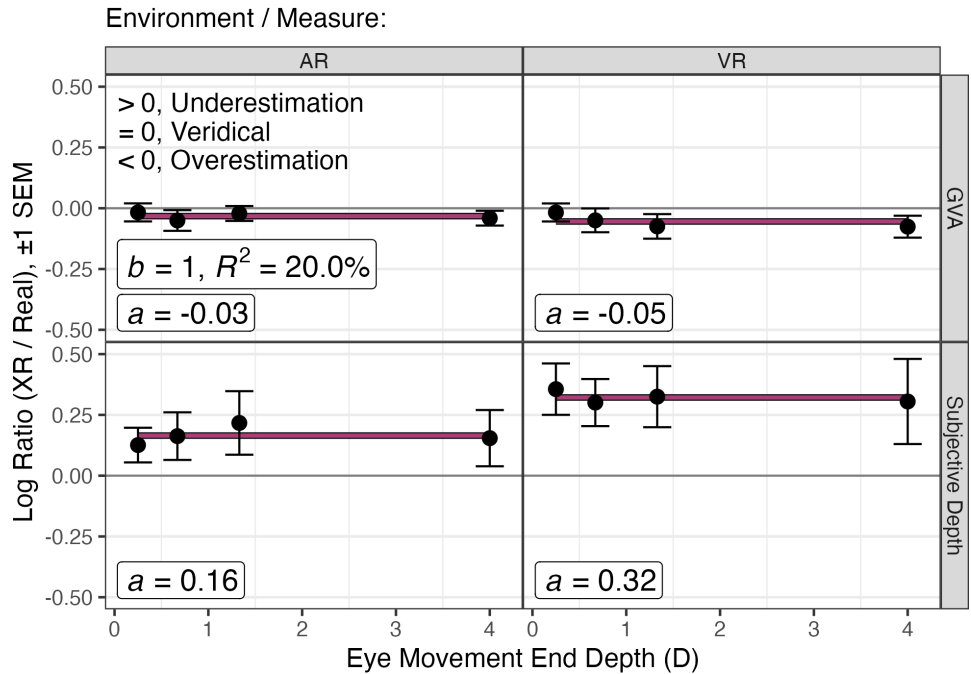

Figure 12: The log ratio of *XR over real* (log( *XR / real* )), predicted by eye movement *end depth*, *environment* (AR, VR) and *measure* (GVA, subjective depth). The fitted linear model *log ratio ~ environment * measure* is shown. Overall, subjective depth judgements were underestimated, consistent with prior findings. However, GVA measurements were much closer to unity, showing a small degree of overestimation. $N = 208$ data points.

```
Model   Formula                              R2     Res.Df  Df    F      p
cm      LogRatio ~ EndDepth * Env * Measure  20.10%  200
fm      LogRatio ~ Env * Measure             20.00%  204    -4   0.1    0.993
rm      LogRatio ~ Measure                   17.36%  206    -2   3.3   <0.050 *

R2 explained by:
      Env = (fm-rm)/fm = 13.2%
  Measure =      rm/cm = 86.4%
EndDepth = (cm-fm)/cm =  0.5%
```

Figure 13: Multiple regression analysis from Figure 12. The fitted model is highlighted. See the caption for Figure 5.

Table 2: Correlation between *subjective depth* in diopters and *normalized GVA*.

| Environment | Correlation *r* | Correlation $R^2$ |
|---|---|---|
| real | .624 | 39.0% |
| AR | .762 | 58.1% |
| VR | .506 | 25.6% |

Finally, we hypothesized that subjective depth judgments would be underestimated, and that GVA would provide a more veridical depth estimate than subjective depth judgements (H4). To begin this analysis, it was first necessary to average both verbal reports and GVA measurements into the same count of measurements. The verbal depth reports, which were given in the unit of measurement preferred by the participant, were converted to meters and then into diopters: *subjective depth in diopters = 1 / verbally reported depth in meters*. A total of 13 (participant) * 3 (environment) * 4 (end depth) * 3 (repetition) = 468 verbal reports (trials) were collected, and then averaged across repetitions into *N* = 156 subjective depth judgements. These judgements were then compared to the 2489 vergence measurements, which (as above) were also averaged into 13 (participant) * 3 (environment) * 4 (end depth) = 156 GVA values. Although the two *measures* (GVA, subjective depth) are expressed in different units (degrees, diopters) and likely come from different distributions, we expected a high degree of correlation. Because normalized GVA removed most between-subject differences, the correlation between normalized GVA and subjective depth was examined. Table 2 shows the Pearson's correlation coefficient for each environment. The correlations, *r*, are positive, indicating that increasing normalized GVA is associated with increasing subjective depth in diopters (i.e. decreasing subjective depth in meters), both in agreement with the geometry of binocular vision (Figure 1). Examining the correlations as $R^2$ indicates that normalized GVA explains a reasonable percentage of the variation in subjective depth judgments.

Next, we sought a way of directly comparing subjective depth and GVA measurements. As shown in Figure 12, for each participant we calculated the log ratio (*XR / Real*) for both GVA and subjective depth measurements. This ratio considers responses to real targets to represent ground truth, and effectively represents both GVA and subjective depth on the same unitless ratio scale. Here, a *y*-axis value near 0 indicates a veridical response with respect to the depth of the real target; values > 0 indicate underestimated depths and values < 0 indicate overestimated depths. The natural logarithm transform normalizes the distribution of ratios to be symmetric around 0 and is more suitable for standard statistical analyses (Bland & Altman, 1996). Figure 12 shows the effects of *environment* (AR, VR), *measure* (GVA, subjective depth), and the eye movement *end depth* on log ratio. A complete linear model was created (Figure 13) that predicts log ratio from the contributions of each predictor: *log ratio ~ end depth * environment * measure*. This complete model explains $R^2$ = 20.1% of the variation. As shown in Figure 12, the fitted model is *log ratio ~ environment * measure*, suggesting that log ratio is significantly predicted by the interaction of environment and measure. This fitted model explains $R^2$ = 20.0% of the variation, an insignificant difference of $dR^2$ = .1% ($F_{4, 200}$ < 1). The fitted model explains significantly more variation ($F_{2, 204}$ = 3.3, $p$ < .05) than the next reduced model, *log ratio ~ measure*. Of the explained variation, 86.4% (of 20.1%) is explained by measure, while 13.2% (of 20.0%) is explained by environment. These differences can be seen in Figure 12, where measure is clearly the strongest effect. Notable here is just how little explanation comes from end depth: 0.5% (of 20.1%).

This analysis and fitted model shows that, consistent with hypothesis H4, subjective depth judgements were underestimated. Turning the log ratios back into ratios, AR targets were underestimated by a factor of 17% ($e^{0.16}$ = 1.17), and VR targets by a factor of 38% ($e^{0.32}$ = 1.377). Also consistent with H4, GVA provides a more veridical depth estimate than subjective depth judgements. The GVA response to AR targets was overestimated by a factor of just 3% ($e^{-0.03}$ = .9704), and VR targets by a factor of 4.9% ($e^{-0.05}$ = .9512). As suggested by the straight lines, and the very small end depth explanation of 0.5%, these effects were constant across the tested distances.

# Discussion

The main goal of this research was to understand how eye vergence angle (GVA) mapped onto depth perception of real and virtual objects in a range of extended reality environments. Here, participants donned an eye tracker and HoloLens 2 while they fixated on real, AR, and VR targets located at four different depth positions (.25, .67, 1.33, and 4 diopters). We hypothesized that GVA would co-vary with the fixated depth (H1), but this covariation would vary by environment (H2). We also hypothesized that GVA would be stable at the fixated depth regardless of the amount of the switching depth and eye movement direction (H3). With regards to subjective depth measurements, we hypothesized that subjective reports would underestimate the depth of XR objects by comparison to real objects, and that objective GVA measurements would provide a more veridical estimate of depth (H4).

## End Depth (H1)

Our first hypothesis (H1) stated that vergence angle would co-vary with fixed target depth. GVA was found to be consistent with the geometry of perceived depth (Figure 1): GVA was larger for near targets and smaller for far targets, and this pattern followed the expected non-linear relationship that relates vergence angle to depth (Figure 4a) as reported in previous studies (Arefin, Swan, et al., 2022; Iskander et al., 2019). By performing a non-linear transform of distance and converting it to diopters (1/distance), we observed a strong linear relationship between GVA and target depth in diopters (Figure 4b). This transformation of units to diopters allowed us to employ standard statistical modeling approaches that assume such linear relationships. Our results strongly support H1 and the validity of our algorithm for accurately measuring GVA from eye-tracking data using the Pupil Labs Pupil Core device.

An interesting and unexpected finding was a strong per-participant GVA bias that was constant across the range of target distances. In the experiment, this bias took on the form of a per-participant intercept (Figure 6). This raises an important question: Which factors contributed to this large bias?

One possible explanation for the per-participant bias is that it could result from variation in the alignment of the eye tracking cameras with the eyes. Anecdotal evidence for this explanation is that the experimenters noticed that, when measuring their own GVA, repeated measurements that involved removing and then putting the AR display back on their heads and recalibration resulted in baseline shifts demonstrating similarly large differences in intercept (but not slope). For a given observer, each time the AR display is placed back on the head, the calibration, position, and orientation of the eye tracking cameras relative to the eyes changes, which could result in this kind of constant GVA bias that shows up as an intercept. Although this explanation was not examined systematically here, similar effects have been found in AR calibration studies, which are also sensitive to small changes in the way that an AR display sits on the head (e.g., Moser et al., 2015).

This suggests a useful future experiment, in which a participant systematically removes and dons the headset while collecting similar GVA measurements. Furthermore, in our study, we mounted a Pupil Labs Pupil Core eye tracker on a Microsoft HoloLens 2, because at the time the Microsoft eye tracking API did not provide separate gaze vectors for the left and right eyes. However, Microsoft recently added this capability to their eye tracking API, so a future study could also replicate the experiment with the HoloLens 2 integrated eye tracking cameras and might expect a reduction in the per-participant bias.

Another possible explanation for the per-participant bias is interpupillary distance, where larger distances result in larger vergence angles, and smaller distances result in smaller vergence angles. To test this idea, the interpupillary distances of the 13 participants were regressed against the intercepts shown in Figure 6 (*intercept ~ interpupillary distance*). This regression did not predict intercept ($R^2$ = 1.4%, $F_{1, 11}$ < 1), and so the current experiment did not find evidence for this explanation. However, an important caveat is that 13 participants, each contributing a single data point, is too small of a sample size to disprove this hypothesis. In addition, interpupillary distance was measured at infinity, not at the distances tested in the experiment. Therefore, a future study could replicate the experimental procedure while measuring interpupillary distance at each target distance, and again examine any per-participant bias in GVA.

## Environment (H2)

Our second hypothesis (H2) was that vergence angle would differ according to whether targets were in a real, AR, or VR environment. H2 was not supported when GVA was examined (Figure 4b), but when the large per-participant intercepts were removed to produce normalized GVA measurements, H2 was supported (Figures 7b, 9). Nevertheless, the effect of the environment was relatively small: GVA measurements for AR targets were .8° smaller than for real targets, and VR targets were 1.3° smaller than for real targets. In terms of the variance explained by the different models, the environment explained only 4.2% (Figure 5), 3.8% (Figure 8), and 5.0% (Figure 10). In addition, the models did not find any interaction effects involving the environment, meaning the effect was constant across the tested ranges of the target depth and focal switching depth.

Whether or not this environmental effect is of practical importance will depend upon the application, and the degree to which GVA predicts perceptual performance. The effect could be significant for tasks performed at reaching distances. The experiment showed that a change in GVA as large as 1.3° (Figure 7b) could result in responses to XR targets that were overestimated between 3% (AR) to 4.9% (VR) (Figure 12). For a numerical example, at a reaching distance of 40 cm, these could result in a depth change of 1.2 cm (AR) to 2 cm (VR). Compare this to the recommendation from Edwards et al. (2004), who from clinical experiments suggested that for AR to be useful in image-guided brain surgery, depth error tolerances of 1 mm or less would be required. On the other hand, for tasks performed at action space distances of about 1.5 to 30 meters (Cutting, 1997), or for tasks where GVA is not strongly related to perceptual performance, the environmental effect may be less practically important.

One possible explanation for this environmental effect on GVA is the vergence accommodation conflict (VAC). Previous research has observed the after-effect of the VAC on the dynamics of the vergence mechanism (Vienne et al., 2014). The VAC also causes the vergence angle to bias toward the depth of a display's fixed focal plane, which can bias perceived depth (Kruijff et al., 2010; Singh, Ellis, & Swan, 2018). Our experiment used the Microsoft HoloLens 2, which has a fixed focal plane at a depth of approximately 1.5 meters (0.67 D). As the real target objects were displayed on monitors, the measured GVA results were not affected by the VAC. However, AR and VR targets should be affected. For the AR and VR targets at .75 and .25 meters (1.33 and 4 D), located in front of the fixed focal plane, GVA values should be biased by the VAC to be smaller than they otherwise would be, corresponding to depths that are farther than they otherwise would be. Figure 7b supports this hypothesis, where at 1.33 and 4 diopters GVA values for AR and VR targets are smaller than real targets. However, this hypothesis also predicts that for the AR and VR targets at 4 meters (.25 D), located behind the fixed focal plane, GVA values should be biased to be larger than they otherwise would be, corresponding to depths that are closer than they otherwise would be. Figure 7b does not support this hypothesis, which found the GVA bias to be constant over all tested distances. Therefore, while the experiment found that VAC could explain some of the effect of environment, it could not explain all of it.

The brightness of an object is a perceptual attribute that is driven by a combination of object luminance and the contrast between the object and its background. Although in the current study, luminance and brightness were not measured, it is likely that participants experienced the VR targets as brightest, followed by the AR targets, and then experienced the real targets as dimmest. This is because the VR targets were rendered by a near-eye display against a completely black background, the AR targets were rendered by the same display against a lit background, and the real targets were rendered on monitors positioned in the background. Huckauf (2018) investigated how brightness during eye tracker calibration and brightness during a task affect EVA change for a single depth at 63 cm. When they performed the calibration on a bright background and considered the task on a dark background, the vergence angle became smaller and moved further away from the observer. In our experiment, the eye tracker calibration was conducted on a bright background (the Pupil Labs screen-based calibration), but the experimental stimuli were presented on a dark background, with the maximum possible contrast for the VR targets. Similar to the findings of Huckauf (2018), the largest EVA was found for the real targets, followed by the AR targets, and then the VR targets had the smallest EVA. Therefore, brightness and contrast effects described by Huckauf (2018) are consistent with the environment effects found here.

This possible reason suggests a future study in which GVA is again measured in real, AR, and VR environments, while luminance is measured and systematically manipulated, along with a perceptual measure of perceived brightness. The future study would be further enriched by also measuring perceived depth using a technique such as disparity matching (e.g., Singh et al., 2018). Such an experiment could examine how GVA behaves in the presence of perceived depth in many conditions and address many unanswered questions about the perception of virtual and real objects.

## Vergence Stability (H3)

We hypothesized that vergence angle would be stable regardless of the focal switching depth and vergence direction (H3). We tested this hypothesis in real, AR, and VR environments (Figure 9).  In this analysis, the vergence direction of the eye movement was not an independent variable, but varied according to focal switching depth. Note that all eye movements diverged for the farthest distance of .25 D, while all eye movements converged for the closest distance of 4 D. Depending on the location of the prior distance, eye movements could converge or diverge at the intermediate distances of .67 and 1.33 D. The linear modeling (Figure 9) shows a strong effect of end depth (94.2% of explained variation), a small but detectable effect of environment (5.0%), and no effect of focal switching depth. As with H2, the effect of environment was not detected when GVA was examined (Figure 11), but when the large per-participant intercepts were removed to produce normalized GVA measurements, the effect of environment was detected (Figure 10).  Notable in the linear modeling is the very small effect size of switching depth: 0.9% of explained variation. Taken together, these effect sizes are statistical evidence for the vergence stability hypothesis, supporting hypothesis H3. This result indicates that GVA does not specifically depend on the start depth and vergence direction: no matter the size of the vergence eye movement, a stable vergence angle associated with the target's depth location was measured. A possible reason for this finding is the vergence system's components in response to depth changes.  The vergence system can quickly detect the rapid depth change with the motor signal (the fast vergence component) and can receive continuous neurological feedback for vergence correction through the oculomotor response (the slow feedback component) (Hung et al., 1996; Vienne et al., 2014).

## Subjective Depth Judgments (H4)

We used subjective measurement in the form of verbal reports to obtain participants' perceptual experience of depth for each target location. We hypothesized (H4) that subjective depth judgements would show depth underestimation for virtual targets, consistent with prior research (Creem-Regehr et al., 2015; Gagnon et al., 2021; Interrante et al., 2006; Kelly et al., 2018; Swan et al., 2007, 2015), and that objective measurements of vergence angle would provide a more veridical estimate of depth (neither underestimation or overestimation). Overall, the results supported this hypothesis. As shown in Figure 12, we observed a significant amount of underestimation from subjective depth reports for virtual target depth in AR and VR environments compared to real targets. This effect of underestimation was constant across depth locations and was largest for VR objects (38%) compared to AR objects (17%). We observed a slight overestimation for objective GVA measures of depth in AR (3%) and VR (4.9%), but this differential was much closer to the veridical depth of real targets compared to the subjective reports.

Critically, these results suggest that the oculomotor system of users in AR and VR produces vergence eye movements that nearly reflect the veridical distance of virtual targets, as evidenced by GVA measurements that were nearly identical to the vergence eye movements induced by real targets at the same distances. This result casts new light on much previous work that reported that subjective measurements of perceived depth tend to underestimate the actual depth compared to real targets (Gagnon et al., 2021; Interrante et al., 2006; Kelly et al., 2018). Our work suggests that the vergence system itself is likely not the cause of this subjective bias; thus, the tendency for subjective depth underestimation for virtual targets must arise from downstream effects such as cognitive biases in reporting or perceptual illusory biases introduced at a later stage of visual information processing. While our study was not designed to elucidate the nature of such downstream effects on subjective depth perception (whether perceptual, cognitive or both), future studies could be designed with the aim of disentangling these factors. We believe that such future studies would benefit from simultaneous use of eye-tracking to objectively measure GVA to help establish a ground truth for the depth of fixated objects during the task and rule out eye vergence behavior as a potential explanation for observed effects on subjective reports of perceptual depth. Therefore, a potential future study should consider a *definite distance perception* (e.g., Bingham & Pagano (1998) mechanism, including both verbal reports and depth matching between real and virtual objects while measuring GVA.

# Conclusion

This research reports one of the first experiments to measure gaze-measured eye vergence angle (GVA) systematically in real and carefully calibrated extended reality (XR) environments, considering four different depth locations: .25, .75, 1.5, and 4 meters (4, 1.33, .67, .25 diopters). Overall, the main findings are:

**GVA, Fixated Depth, Individual Differences**: GVA was strongly associated with the fixated depth in the expected non-linear pattern that relates vergence angle to depth. By performing a non-linear transform of distance and converting it to diopters (1/distance), a strong linear relationship was observed between GVA and target depth in diopters. However, large individual variations in GVA were observed, which may arise from eye tracking calibration, eye camera position, interpupillary distance, brightness and contrast effects, or other unknown factors.

**GVA and XR Environment**: When the large individual variations in GVA were removed, GVA differed according to whether the targets were in a real, augmented reality (AR), or virtual reality (VR) environment. The detected differences were small: GVA measurements for AR targets were .8° smaller than for real targets, and VR targets were 1.3° smaller than for real targets. Whether or not these environmental effects will be practically important will depend upon the application, and the unknown degree to which GVA predicts perceptual performance. The environment effects may arise from the vergence accommodation conflict, brightness and contrast effects, or other unknown factors.

**Vergence Stability**: Measured GVA was stable with respect to the starting depth of the previously fixated target and invariant to the direction (convergence, divergence) of the vergence eye movement. This implies that there is an unbiased one-to-one mapping between GVA and the depth of real and virtual targets.

**GVA and Subjective Depth**: Subjective depth from verbal reports underestimated the depth of XR targets compared to real targets, while GVA showed very little bias for XR targets compared to real targets. Therefore, a novel finding from this research is that GVA provided a much more veridical estimate of AR and VR target depth, compared to subjective judgment.

Recently, to provide improved and automated calibration, many head-mounted XR displays (e.g., Microsoft HoloLens 2, Magic Leap 2, HTV Vive Pro Eye, and others) have included built-in eye tracking. In addition to calibration, eye tracking could enable novel ways of interacting with XR content. Furthermore, such systems will allow investigating novel research questions about how the human visual system operates when engaged by virtual objects at different depths.

# Acknowledgments

Research was sponsored by the Army Research Laboratory and was accomplished under Cooperative Agreement Number W911NF-22-2-0227. The views and conclusions contained in this document are those of the authors and should not be interpreted as representing the official policies, either expressed or implied, of the Army Research Laboratory or the U.S. Government. The U.S. Government is authorized to reproduce and distribute reprints for Government purposes notwithstanding any copyright notation herein. We acknowledge the Center for Advanced Vehicular Systems (CAVS) at Mississippi State University for providing the laboratory space where this experiment was conducted.